\documentstyle[aps,12pt]{revtex}
\newcommand{\bb}{\begin{equation}}
\newcommand{\ee}{\end{equation}}
\newcommand{\ba}{\begin{array}}
\newcommand{\ea}{\end{array}}

\begin {document}

\title{Dirac particles in a rotating magnetic field\thanks
{published in J. Phys. A {\bf 34} (2001) 1903-1909.}}
\author{Qiong-gui Lin\thanks{E-mail addresses:
        qg\_lin@163.net, qg\_lin@263.net}}
\address{China Center of Advanced Science and Technology (World
    Laboratory),\\
        P.O.Box 8730, Beijing 100080, People's Republic of China
        \thanks{not for correspondence}\\
        and\\
        Department of Physics, Zhongshan University, Guangzhou
        510275,\\
        People's  Republic of China}

\maketitle
\vfill

\begin{abstract}
\baselineskip 15pt{\normalsize We study a relativistic charged Dirac
particle moving in a rotating magnetic field. By using a
time-dependent unitary transformation, the Dirac equation with the
time-dependent Hamiltonian can be reduced to a Dirac-like equation
with a time-independent effective Hamiltonian. Eigenstates of the
effective Hamiltonian correspond to cyclic solutions of the original
Dirac equation. The nonadiabatic geometric phase of a cyclic solution
can be expressed in terms of the expectation value of the component
of the total angular momentum along the rotating axis, regardless of
whether the solution is explictly available. For a slowly rotating
magnetic field, the eigenvalue problem of the effective Hamiltonian
is solved approximately and the geometric phases are calculated. The
same problem for a charged or neutral Dirac particle with an
anomalous magnetic moment is discussed briefly.}
\end{abstract}
\vfill
\leftline {PACS number(s): 03.65.Pm, 03.65.Vf}
\newpage
\baselineskip 15pt

In quantum mechanics, the Schr\"odinger equation, even with a
time-independent Hamiltonian, can be solved analytically only in a
few cases. With a time-dependent Hamiltonian, the problem is more
difficult and fewer examples are well studied. One of the well
studied examples is a nonrelativistic neutral particle with spin and
magnetic moment in a rotating magnetic field [1-5]. This simple
example has received much attention because of its relevance to the
problem of geometric phases [6-8], and also because exact solutions
are available. In a recent work [9] we have studied a nonrelativistic
charged particle moving in a rotating magnetic field, with or without
a central potential. The case with a central potential can describe
the valence eletron of an alkaline atom or that of a hydrogen atom
under the influence of the external magnetic field. The problem can
be treated analytically and some exact solutions are available. The
purpose of this paper is to extend the previous work to a
relativistic charged Dirac particle (without a central potential).
The main interest of the problem is to see how the previous results
are changed by the relativistic effect.

Consider a charged particle with spin $1/2$, moving in a
rotating magnetic field. The motion is described by the Dirac
equation. This equation can be written in a form similar to the
Schr\"odinger one. Thus the ideas of cyclic motion,
nonadiabatic geometric phase, etc. for the Schr\"odinger equation [7]
also apply here. The Hamiltonian for this system is of course
time dependent. As in the nonrelativistic case, we use a
time-dependent unitary transformation to reduce the Dirac equation to
a Dirac-like one with a time-independent effective Hamiltonian.
This is equivalent to transforming the equation into a rotating frame
[10] where the magnetic field is static. Thus the effective
Hamiltonian in that frame is time independent.
However, as emphasized in Ref. [9], the transformation is merely
a mathematical technique, and the effective equation in the rotating
frame (which is not an inertial system) does not describe a real
physical problem. The results derived from this equation are not
observable in the rotating frame.

As in the nonrelativistic case, it can
be shown that eigenstates of the effective Hamiltonian correspond to
cyclic solutions of the original Dirac equation. The
nonadiabatic geometric phase of a cyclic solution can be expressed in
terms of the expectation value of the component of the total angular
momentum along the rotating axis, regardless of whether the solution
is explictly available. For a slowly rotating magnetic field, some of
the terms in the effective Hamiltonian can be treated as small
perturbations and the engenvalue problem of the remaining terms can be
solved exactly. In this approximation, the geometric phases of the
cyclic solutions can be calculated explicitly.
The above described procedure can also be applied to the Dirac-Pauli
equation [11] for a charged or neutral particle with an anomalous
magnetic moment, except that the effective Hamiltonian is too
complicated and its eigenvalue problem has not been solved.

We begin with the Dirac equation
\bb
\left[i\gamma^\mu\left(\partial_\mu+{iq\over\hbar c}A_\mu\right)-
{Mc\over \hbar}\right]\Psi=0,
\ee     
where $M$ and $q$ are respectively the mass and electric charge of the
particle, $A_\mu$ is the vector potential describing the external
electromagnetic fields. The latter is chosen as
\bb
A_0=0,\quad
{\bf A}(t)=\textstyle{\frac 12}{\bf B}(t)\times{\bf r}.
\ee     
This vector potential produces the magnetic field ${\bf B}(t)$ which
is chosen to be one with a constant magnitude $B$ and rotating about
some fixed axis at a constant angle $\theta_B$ and with a constant
frequency $\omega$. The rotating axis is chosen as the $z$ axis of
the coordinate system. The magnetic field is therefore
\bb {\bf
B}(t)=B{\bf n}(t), \quad {\bf n}(t) =(\sin\theta_B\cos\omega t,
\sin\theta_B\sin\omega t, \cos\theta_B).
\ee     
We take $B>0$ without loss of generality. Note that ${\bf A}(t)$ also
generates a time-varying electric field. Thus we are indeed dealing
with a time-varying electromagnetic field. However, the electric
field does not enter the Dirac equation directly if the particle has
no anomalous magnetic moment. We write the Dirac equation in the
following form:
$$
i\hbar\partial_t\Psi=H(t)\Psi,
\eqno(4{\rm a})$$
where
$$
H(t)=c{\bbox\alpha}\cdot\left[{\bf p}-{q\over c}{\bf A}(t)\right]
+Mc^2\gamma^0,
\eqno(4{\rm b})$$
where ${\bbox\alpha}=\gamma^0{\bbox\gamma}$. To solve the equation
we define the orbit angular momentum (in unit of $\hbar$)
${\bf l}={\bf r}\times{\bf p}/\hbar$, the spin angular momentum
${\bf s}={\bbox\Sigma}/2$ where $\Sigma^i=i\epsilon^{ijk}\gamma^j
\gamma^k/2$, and the total angular momentum
${\bf j=l+s}$; and then make a time-dependent unitary transformation
\addtocounter{equation}{1}
\bb
\Psi(t)=W(t)\Phi(t),
\ee     
where
\bb
W(t)=\exp(-i\omega t j_z),
\ee     
and $j_z$ is the $z$-component of the total angular momentum ${\bf
j}$. This transformation is a generalization of that used in solving
the Schr\"odinger equation for a neutral [2] or charged [9] particle
with spin in the rotating magnetic field. It is not difficult to show
that \bb W^\dagger(t)H(t)W(t)=H(0).
\ee     
Thus we obtain the following equation for $\Phi$:
\bb
i\hbar\partial_t\Phi=H_{\rm eff}\Phi,
\ee     
where the effective Hamiltonian
\bb
H_{\rm eff}=H(0)-\hbar\omega j_z.
\ee     
Since $H_{\rm eff}$ is time-independent, Eq. (8) has the formal
solution
\bb
\Phi(t)=U_{\rm eff}(t)\Phi(0),\quad
U_{\rm eff}(t)=\exp(-iH_{\rm eff}t/\hbar).
\ee     
With the obvious relation $\Psi(0)=\Phi(0)$, the time-dependent
Dirac equation (4) is formally solved as
\bb
\Psi(t)=U(t)\Psi(0),\quad
U(t)=W(t)U_{\rm eff}(t).
\ee     
Since $U(t)$ involves no chronological product, this solution is
convenient for practical calculations.

Now we show that eigenstates of the effective Hamiltonian correspond
to cyclic solutions of Eq. (4). We take the initial condition
\bb
\Psi_i(0)=\varphi_i,
\ee     
where $\varphi_i$ is an eigenstate of the effective Hamiltonian with
eigenvalue $E_i$,
and calculate $\Psi_i(T)$ where $T=2\pi/\omega$ is the period of the
rotating magnetic field. Here for convenience we use one subscript $i$
to represent all the quantum numbers that is needed to specify an
eigenstate. Obviously,
$U_{\rm eff}(t)\varphi_i=\exp(-iE_{i}t/\hbar)\varphi_i$, valid for all
$t$, and $W(T)\varphi_i=\exp(-i2\pi j_z)\varphi_i$. Because
we can always expand $\varphi_i$ as a
linear combination of the eigenstates of $j_z$, we obtain
\bb
\Psi_i(T)=\exp(-iE_{i}T/\hbar-i\pi)\Psi_i(0).
\ee     
Hence it is indeed a cyclic solution, and the total phase change
in a period is
\bb
\delta_i=-E_{i}T/\hbar-\pi, \quad {\rm mod} ~2\pi.
\ee     
To determine the dynamic phase, we should calculate
$$
\langle H(t)\rangle_i\equiv (\Psi_i(t), H(t)\Psi_i(t))
=(\Psi_i(0),W^\dagger H(t)W\Psi_i(0))
=(\varphi_i, H(0)\varphi_i).
$$
Because $H(0)=H_{\rm eff}+\hbar\omega j_z$, we have
\bb
\langle H(t)\rangle_i=E_i+\hbar\omega\langle j_z\rangle_i.
\ee     
Here $\langle j_z\rangle_i=(\varphi_i, j_z\varphi_i)=
(\Psi_i(t), j_z\Psi_i(t))$ is the expectation value of $j_z$ in the
state $\Psi_i(t)$, and it is time independent. Note that
$\langle H(t)\rangle_i$ is also independent of $t$. Thus the state
$\Psi_i(t)$ is somewhat similar to a stationary state in a system
with a time-independent Hamiltonian. The dynamic phase is
\bb
\beta_i=-\hbar^{-1}\int_0^T dt\;\langle H(t)\rangle_i=
-E_iT/\hbar-2\pi\langle j_z\rangle_i.
\ee     
Therefore the nonadiabatic geometric phase is
\bb
\gamma_i=\delta_i-\beta_i=-\pi+2\pi\langle j_z\rangle_i,
\quad {\rm mod} ~2\pi,
\ee     
and is determined by $\langle j_z\rangle_i$. This is a relativistic
generalization of the result for a nonrelativistic neutral [2] or
charged particle [9], and has the same form as the corresponding
nonrelativistic result. It is valid regardless of whether $\varphi_i$
is explicitly available or not, and
is convenient for approximate calculations if necessary.

Our next task is to find the eigenvalues and eigenstates of $H_{\rm
eff}$. Since the effective Hamiltonian is somewhat complicated, we
have to make some approximation. We assume that $\omega$ is small
such that the term $-\hbar\omega j_z$ in $H_{\rm eff}$ can be treated
as a small perturbation. That is, we are considering a slowly
rotating magnetic field. In the nonrelativistic case [9], the
restriction is specifically $\omega\ll |q|B/2Mc$, and we have argued
that this is in fact a rather loose restriction. From the following
result for the energy levels we would see that the argument also
holds in the relativistic case. We thus decompose $H_{\rm eff}$ as
\bb H_{\rm eff}=H^0_{\rm eff}+H'_{\rm eff},
\ee     
where
\bb
H^0_{\rm eff}=c{\bbox\alpha}\cdot\left({\bf p}-{q\over c}{\bf A}_0
\right)+Mc^2\gamma^0,
\ee     
whoes eigenvalue problem will be solved exactly, and
\bb
H'_{\rm eff}=-\hbar\omega j_z
\ee     
which will be treated as a small perturbation. In Eq. (19)
${\bf A}_0={\bf A}(0)$. Note that $H^0_{\rm eff}=H(0)$.
It is not difficult to show that
\bb
H^0_{\rm eff}=\exp(-i\theta_B j_y)H^z_{\rm eff}
\exp(i\theta_B j_y),
\ee     
where
\bb
H^z_{\rm eff}
=c{\bbox\alpha}\cdot\left({\bf p}-{q\over c}{\bf A}_z
\right)+Mc^2\gamma^0,
\ee     
${\bf A}_z=\frac 12{\bf B}_z\times {\bf r}$ and ${\bf B}_z=B{\bf
n}_z=B(0,0,1)$. This is the Hamiltonian of a relativistic charged
particle in a static uniform magnetic field along the $z$ axis.

The eigenvalue problem of $H^z_{\rm eff}$ is easy.
We write down the eigenvalue equation
\bb
H^z_{\rm eff}\zeta=E^0\zeta
\ee     
and denote $\zeta=(u,v)^\tau$ where $u$ and $v$ are two-component
spinors and the superscript $\tau$ denotes matrix transposition. In
terms of the two-component spinors the above equation takes the form
\begin{eqnarray}
&& c{\bbox\sigma}\cdot\left({\bf p}-{q\over c}{\bf A}_z\right)u
=(E^0+Mc^2)v,\nonumber\\
&& c{\bbox\sigma}\cdot\left({\bf p}-{q\over c}{\bf A}_z\right)v
=(E^0-Mc^2)u.
\end{eqnarray}  
One can solve the first for $v$, and substitute it into the second.
Then an equation in $u$ alone is obtained. It can be solved
in the cylindrical coordinates $(\rho,\phi, z)$.
The energy eigenvalues are
\bb
E^0_i=E^0_{n_zn_{\rho}mm_s\pm}=\pm\{(Mc^2)^2+(\hbar ck_z)^2+
|q|B\hbar c[2n_\rho+|m|+1-\epsilon(q)(m+2m_s)]\}^{1/2},
\ee     
where we use a single subscript $i$ to represent all the quantum
numbers and the sign of the energy; $k_z=2\pi n_z/d$ where
$d$ is a length in the $z$ direction for box normalization and
$n_z=0,\pm1,\pm2,\ldots$; $n_\rho=0,1,2,\ldots$ is a radial quantum
number; $m=0,\pm1,\pm2,\ldots$ and $m_s=\pm 1/2$; $\epsilon(q)$
is a sign function of $q$. The corresponding eigenfunctions are
given by
\bb
u_i(\rho,\phi, z, s_z)=N_{i}e^{-\alpha^2\rho^2/2}(\alpha\rho)^{|m|}
L_{n_\rho}^{|m|}(\alpha^2\rho^2){e^{im\phi}\over\sqrt{2\pi}}
{e^{i k_z z}\over \sqrt d}\chi_{m_s}(s_z),
\ee     
where $\alpha=\sqrt{|q|B/2\hbar c}$, the $L_{n_{\rho}}^{|m|}$ are
Laguerre polynomials [12], $\chi_{m_s}(s_z)$ is the
eigenstate of $s_z$ with eigenvalue $m_s$; and
\bb
v_i(\rho,\phi, z, s_z)={c\over E^0_i+Mc^2}
{\bbox\sigma}\cdot\left({\bf p}-{q\over c}{\bf A}_z\right)
u_i(\rho,\phi, z, s_z).
\ee     
We do not write down the specific form of $v_i$ because it is lengthy
and not necessary for the subsequent calculations. We just mention
that it consists of two terms. One is proportional to $u_i$, and the
other involves the factor $e^{i(m-1)\phi}\chi_{m_s+1}(s_z)$ or
$e^{i(m+1)\phi}\chi_{m_s-1}(s_z)$, depending on whether $m_s=-1/2$ or
$m_s=1/2$, respectively. It should be remarked
that neither $m$ nor $m_s$ is a good quantum number. A good quantum
number related to them is $m_j=m+m_s$ which is the eigenvalue of
$j_z$, a conserved quantity. The reason why we need the
quantum number $m_s$ is that for a given $m_j$ there are two kinds
of solutions, corresponding to the two values of $m_s$. In the
above solutions the normalization constants are
\bb
N_{i}=\alpha\left({E_i^0+Mc^2\over E_i^0}\right)^{1/2}
\left[{n_{\rho}!\over \Gamma(n_{\rho}+|m|+1)}\right]^{1/2}.
\ee     
Note that the solutions in the above forms are not appropriate when
$E_i^0=-Mc^2$, which may happen when $n_z=0$, $n_\rho=0$,
$m=\epsilon(q)|m|$, $m_s=\epsilon(q)|m_s|$. Indeed, one cannot
eliminate $v$ from Eq. (24) in this case. Rather, one should
eliminate $u$ and solve the resulted equation for $v$. The solution
reads
\bb v_{00mm_s-}(\rho,\phi, z,
s_z)=N'_{00mm_s-}e^{-\alpha^2\rho^2/2}
(\alpha\rho)^{|m|}{e^{im\phi}\over\sqrt{2\pi}}{1\over \sqrt d}
\chi_{m_s}(s_z),
\ee     
\bb
u_{00mm_s-}(\rho,\phi, z, s_z)=-{1\over 2Mc}
{\bbox\sigma}\cdot\left({\bf p}-{q\over c}{\bf A}_z\right)
v_{00mm_s-}(\rho,\phi, z, s_z).
\ee     
We do not write down the specific form for $u_{00mm_s-}$ as before.
The normalization constant in the above solution is
\bb
N'_{00mm_s-}={\sqrt 2 \alpha\over \sqrt{\Gamma(|m|+1)}}.
\ee     
Thus the equation (23) is completely solved. The solutions of this
equation can also be found in Ref. [13], but in different forms.
The reason why the solution with a specific energy level can have
different forms is that the energy levels are degenerate.
It seems that the solutions in our form are more explicit and
convenient.

Now that Eq. (23) is solved, the eigenvalue problem of $H^0_{\rm eff}$
become trivial. The eigenfunctions are
\bb
\varphi_{i}^0=\exp(-i\theta_B j_y)\zeta_i,
\ee     
where $\zeta_i=(u_i,v_i)^\tau$, and the corresponding energy
eigenvalues are still given by Eq. (25). We will use these
$\varphi_{i}^0$ as the approximate eigenfunctions of $H_{\rm eff}$.
Of course the explicit functional form of $\varphi_{i}^0$ is
complicated, but this is not necessary in practical calculations. The
lowest order corrections to the energy eigenvalues are given by the
expectation values of $H'_{\rm eff}$ in the approximate eigenstates.
The corrected energy levels are
\bb
E_i=E_i^0-m_j\hbar\omega\cos\theta_B.
\ee      
Now there is no degeneracy in the quantum numbers.

Our final task is to work out the geometric phase in a period
for the $i$th state, that is, a state with the initial condition (12).
Since we have only an approximate result $\varphi_{i}^0$ for the
eigenstate $\varphi_{i}$, we can calculate the nonadiabatic geometric
phase only approximately. The result is
\bb
\gamma_i=-m_j\Omega_B, \quad {\rm mod} ~2\pi,
\ee     
where $\Omega_B=2\pi(1-\cos\theta_B)$ is the solid angle subtended by
the trace of the rotating magnetic field. For $\omega\ll |q|B/2Mc$,
this has the same value as the corresponding nonrelativistic result
[9]. In the above approximation, it can be shown that
\bb
(\Psi_i(t),
{\bf j}\Psi_i(t))=m_j (\sin\theta_B\cos\omega
t,\sin\theta_B\sin\omega t,\cos\theta_B).
\ee     
Therefore the total angular momentum precesses synchronously with the
magnetic field and approximately at the same angle $\theta_B$
with the rotating axis.
Then $\Omega_B$ is also (approximately) the solid angle subtended by
the trace of the total angular momentum. The geometric nature of the
result (34) is thus obvious. In the nonrelativistic case, both the
orbit and spin angular momentum precess synchronously with the
magnetic field [9]. Here only the total angular momentum does. This
is the main difference caused by the relativistic effect.

Now we turn to the problem of a charged or neutral particle with
anomalous magnetic moment. The Dirac-Pauli equation is
\bb
\left[i\gamma^\mu\left(\partial_\mu+{iq\over\hbar c}A_\mu\right)-
{Mc\over \hbar}-{1\over 2}{\mu_{\rm a}\over\hbar c}\sigma^{\mu\nu}
F_{\mu\nu}\right]\Psi=0,
\ee     
where $\mu_{\rm a}$ is the anomalous magnetic moment,
$F_{\mu\nu}=\partial_\mu A_\nu-\partial_\nu A_\mu$, and
$\sigma^{\mu\nu}=i[\gamma^\mu, \gamma^\nu]/2$. This equation differs
from Eq. (1) by the last term in the square bracket. When $q=0$ it
describes a neutral particle, otherwise it describes a charged one.
With the previous $A_\mu$, it can be recast in the Hamiltonian form:
$$
i\hbar\partial_t\Psi=H(t)\Psi,
\eqno(37{\rm a})$$
where
$$
H(t)=c{\bbox\alpha}\cdot\left[{\bf p}-{q\over c}{\bf A}(t)\right]
+Mc^2\gamma^0-\mu_{\rm a}\gamma^0{\bbox\Sigma}\cdot{\bf B}(t)
+i\mu_{\rm a}{\bbox\gamma}\cdot{\bf E}(t). \eqno(37{\rm b})$$ Note
that the electric field ${\bf E}=-c^{-1}\partial_t {\bf A}$ now
enters the equation directly. We make the time-dependent unitary
transformation (5-6). With some algebra it can be shown that
\addtocounter{equation}{1}
\bb W^\dagger(t)H(t)W(t)=H(0)
\ee     
still holds in the present case. Thus the equation for $\Phi$ has the
same form as given by Eqs. (8-9), and all subsequent discussions
until Eq. (17) are still valid. Unfortunately, $H(0)$ is too
complicated even for a neutral particle and we have not been able to
obtain any eigenstate of it.

In conclusion we have considered a relativistic charged particle
moving in a rotating magnetic field. The Hamiltonian for such a
system is time dependent. By making use of a time-dependent unitary
transformation, the Dirac equation can be reduced to a Dirac-like
equation with an effective Hamiltonian which is time independent. In
this way we obtain a formal solution to the original Dirac equation,
which determines the time evolution of an arbitrary initial state.
The time-evolution operator in this formal solution, unlike that for
a general time-dependent Hamiltonian, involves no chronological
product, and thus is convenient for practical calculations. Any
solution with one of the eigenstates of the effective Hamiltonian as
an initial state is a cyclic solution. The nonadiabatic geometric
phase in a period for such a solution can be expressed in terms of
the expectation value of the component of the total angular momentum
along the rotating axis. This is an exact relation which holds
regardless of whether the solution is explicitly available, and is
convenient for approximate calculations whenever necessary. For a
slowly rotating magnetic field, the eigenvalue problem of the
effective Hamiltonian is solved approximately, and the geometric
phases are calculated. The difference between the relativistic
results and the corresponding nonralativistic ones is discussed. We
also briefly discussed the same problem for a relativistic particle
with an anomalous magnetic moment.

\section*{Acknowledgments}

The author is grateful to Professor Guang-jiong Ni for communications
and encouragement. This work was supported by the
National Natural Science Foundation of China.



\begin{thebibliography}{99}
\baselineskip 15pt

\bibitem{1}L. D. Landau and E. M. Lifshitz, {\it Quantum Mechanics},
           3rd ed. (Pergamon, Oxford, 1977).

\bibitem{2}S.-J. Wang, Phys. Rev. A {\bf 42}, 5107 (1990).

\bibitem{3}A. G. Wagh and V. C. Rakhecha, Phys. Lett. A {\bf 170}, 71
 (1992).

\bibitem{4}G.-J. Ni, S.-Q. Chen, and Y.-L. Shen, Phys. Lett. A
{\bf 197}, 100 (1995).

\bibitem{5}G.-J. Ni and S.-Q. Chen, {\it Advanced Quantum Mechanics}
(Fudan Univ. Press, Shanghai, 2000). (in Chinese)

\bibitem{6}M. V. Berry, Proc. R. Soc. Lond. A {\bf 392}, 45 (1984).

\bibitem{7}Y. Aharonov and J. Anandan, Phys. Rev. Lett. {\bf 58},
1593 (1987).

\bibitem{8}H.-Z. Li, {\it Global Properties of Simple Physical
Systems--Berry's Phase and Others} (Shanghai Scientific \& Technical,
Shanghai, 1998). (in Chinese)

\bibitem{9}Q.-G. Lin, Phys. Rev. A {\bf 63} (2001) 012108.

\bibitem{10}A. G. Wagh and V. C. Rakhecha, Phys. Rev. A {\bf 48},
R1729 (1993).

\bibitem{11}W. Pauli, Rev. Mod. Phys. {\bf 13}, 203 (1941).

\bibitem{12}I. S. Gradshteyn and I. M. Ryzhik, {\it Tables of
Integrals, Series, and Products} (Academic, New York, 1980).

\bibitem{13}V. G. Bagrov and D. M. Gitman, {\it Exact Solutions of
Relativistic Wave Equations} (Kluwer, Dordrecht, 1990).

\end{thebibliography}
\end{document}